\documentclass[11pt]{article} % use larger type; default would be 10pt

\usepackage[utf8]{inputenc} % set input encoding (not needed with XeLaTeX)

\usepackage{geometry} % to change the page dimensions
\usepackage{graphicx} % support the \includegraphics command and options

\usepackage{graphicx}
\usepackage{caption}
\usepackage{color} 

\usepackage{epstopdf}
\usepackage{amsmath} 
\usepackage{animate}
\usepackage{breakcites}
\usepackage{amsfonts}
\usepackage{hyperref}
\usepackage{cite}
\usepackage[affil-it]{authblk}
\usepackage{booktabs} % for much better looking tables
\usepackage{array} % for better arrays (eg matrices) in maths
\usepackage{paralist} % very flexible & customisable lists (eg. enumerate/itemize, etc.)
\usepackage{verbatim} % adds environment for commenting out blocks of text & for better verbatim
\usepackage{subfig}
\usepackage{lipsum}
\usepackage{bm}
\usepackage{fancyhdr} % This should be set AFTER setting up the page geometry
\usepackage{sectsty}
\allsectionsfont{\sffamily\mdseries\upshape} % (See the fntguide.pdf for font help)
\usepackage[nottoc,notlof,notlot]{tocbibind} % Put the bibliography in the ToC
\usepackage[titles,subfigure]{tocloft} % Alter the style of the Table of Contents

\title{Rapid Changes in Synchronizability in Conductance-based Neuronal Networks with Conductance-based Coupling} 

\author[1,2,3*]{Wilten Nicola}

\affil[1]{Hotchkiss Brain Institute}
\affil[2]{University of Calgary, Faculty of Science, Department of Physics and Astronomy}
\affil[3]{University of Calgary, Cumming School of Medicine, Department of Cell Biology and Anatomy}
\affil[*]{%
  Corresponding Author:  wilten.nicola@ucalgary.ca}

\date{\today} % Activate to display a given date or no date (if empty),
\begin{document}
%\linenumbers
\maketitle

\begin{abstract}
Real neurons connect to each other non-randomly.  These connectivity graphs can potentially impact the ability of networks to synchronize, along with the dynamics of neurons and the dynamics of their connections.  How the connectivity of networks of conductance-based neuron models like the classical Hodgkin-Huxley model, or the Morris-Lecar model, impacts synchronizability remains unknown.  One powerful tool to resolve the synchronizability of these networks is the Master Stability Function (MSF).  Here, we apply and extend the MSF approach to networks of Morris-Lecar neurons with conductance-based coupling to determine under which parameters and graphs synchronous solutions are stable.   We consider connectivity graphs with a constant row-sum, where the MSF approach can be readily extended to conductance-based synapses rather than the more well studied diffusive connectivity case, which primarily applies to gap junction connectivity.  In this formulation, the synchronous solution is a single, self-coupled or 'autaptic' neuron.   We find that the primary determining parameter for the stability of the synchronous solution is, unsurprisingly, the reversal potential, as it largely dictates the excitatory/inhibitory potential of a synaptic connection.  However, the change between an ``excitatory" and ``inhibitory'' synapses is rapid, with only a few millivolts separating stability and instability of the synchronous state for most graphs.   We also find that for specific coupling strengths (as measured by the global synaptic conductance), islands of synchronizability in the MSF can emerge for inhibitory connectivity.  We verified the stability of these islands by direct simulation of pairs of neurons coupled with eigenvalues in the matching spectrum.  These results were robust for different transitions to spiking (Hodgkin Class I vs Class II), which displayed very similar synchronizability characteristics.  
\end{abstract}

\section{\textbf{Introduction}}

%introductory paragraph.
Brain cells, like other complex interacting systems can readily synchronize and fire their action potentials or spikes simultaneously, under the right conditions.  Sometimes, this synchronizability is a normal part of brain function.  For example, pyramidal neurons in the hippocampus collectively fire during the ~100-150 millisecond hippocampal sharp-wave ripples, a so-called cognitive biomarker for memory consolidation and memory replay \cite{Buz1,Wilson1,Wilson2}. This synchronization is strong enough to be observed even in the hippocampal local field potential, a macroscopic observable of collective neuronal activity.  However, neurons can also pathologically synchronize \cite{Buz1}.  In fact, the same hippocampal neurons that synchronize in sharp-wave ripples in the hippocampus often synchronize excessively leading to epileptic seizures \cite{Buz1}.    The myriad of conditions that can lead to synchronization or de-synchronization remains an active area of research. 

One hypothesis is that the specific characteristics of the connectivity between neurons can promote or obstruct the ability of neurons to otherwise synchronize.  Thus, with a network that is only constrained by a fixed number of connections, or with a fixed global connection strength, different connectivity profiles can lead to synchronization or any number of asynchronous states. 

Fortunately, there is a tool to analyze how different networks of neuronal models can synchronize: the master stability function (MSF) \cite{msf1,msf2,coombesmsf1,coombesmsf2,coombesmsf3}.  The MSF allows one to determine the stability of a synchronous solution across any connectivity graph through a three-step process. First, the master stability function is computed in the complex plane.   
 Next, the eigenvalues are computed for any particular connectivity graph.  Finally, the sign of the MSF evaluated at the eigenvalues of the graph on the complex plane dictates the local asymptotic stability of the synchronous solution. If all eigenvalues fall in regions where the MSF is negative, the synchronous solution is locally asymptotically stable while a single eigenvalue falling into a region where the MSF is positive destabilizes the synchronous solution  \cite{msf1,msf2}. 

However, the application of the MSF in networks of spiking neurons with conductance-based coupling or chemical synapses has remained under studied for a few reasons.  The first reason is that many neuron models or synaptic models are non-smooth.  For example, all integrate-and-fire neurons utilize a discontinuity to reset the membrane potential of a neuron after a  ``spike", and possibly change other state variables   \cite{Izhikevich1,Izhikevich2,Touboul1,Adex1,Adex2}.  While there have been considerable advances in extending the MSF approach to non-smooth spiking networks and non-smooth differential equations in general \cite{coombesmsf1,coombesmsf2,coombesmsf3,coombesmsf4}, real neurons do not have the types of membrane discontinuities exhibited by integrate-and-fire neurons.   The second reason is that the connections that real neurons form are not entirely ``excitatory" or ``inhibitory" \cite{destexhe}.  The net effect of a chemical synapse depends on the driving force, which is influenced by ionic reversal potentials and the voltage and spike-shape of a neuron itself.  Thus, a connection can switch between excitatory to inhibitory during a single spike depending on the voltage of the neuron at any moment.  Finally, the master stability function is primarily limited to cases where the row-sum of the connectivity matrix is 0, which forces both positive and negative connection weights \cite{msf1,msf2}.  This constraint, which is termed ``diffusive connectivity" is mathematically convenient as it implies that the synchronous solution in the network is also simultaneously a solution to a single uncoupled neuron.  Unfortunately, this constraint is incompatible with chemical synapses where the unitary synaptic conductances, being physical quantities, are non-negative. This is however compatible with gap-junction based connectivity or electrical synapses.  Indeed, this latter case has been extensively studied \cite{gj1,gj2,gj3,gj4,gj5,gj6,gj7}.  

Here, we apply the MSF approach to networks of smooth conductance-based neurons with smooth conductance-based or chemical synaptic coupling.  By utilizing a constant row-sum,  rather than 0 row-sum constraint we successfully applied the MSF approach to analyzing the synchronizability of networks of Morris-Lecar neurons with  chemical synapses \cite{ml1,bardtext}.  This constant row-sum corresponds to a global conductance strength, but not necessarily a measure of inhibition or excitation strength.  We found that independent of the uncoupled neurons' bifurcation to spiking (Hodgkin Class I or Hodgkin Class II), the reversal potential of the synapse would most strongly dictate the stability of synchronous solutions.   We also found that the MSF would rapidly change sign as a function of the reversal potential over the complex plane.  Only a few millivolts separated large-scale stability or instability of the synchronous solution. The global portrait of synchronizability looked similar for both Hodgkin Class I and Class II parameter regimes.  However, the MSF deviated from this result in two ways.  First, the synchronous solution could lose stability depending on the value of the global unitary synaptic conductance and the reversal potential for excitatory coupling parameter ranges.  This loss of stability was readily observed with sufficiently large ring structures. Second, we found that for both classes of firing, islands of stability would emerge for the synchronous solution for inhibitory synapses.   We tested this finding directly with simulations of networks with constrained connectivity matrices that forced the eigenvalues of the matrices into the synchronizability islands.   The synchronizability of any particular graph exhibits regimes that are highly parameter dependent in conductance-based neurons with chemical synapses. Stable configurations can emerge with increasing inhibition and unstable configurations can emerge with increasing excitation.

%paragraph on the different neuron/connectivity types.

%paragraph on the master stability function.

%paragraph on the results.
\section{\textbf{Results}} 
\subsection{\textbf{The Master Stability Function for Conductance Models with Conductance-Based Coupling}}
To investigate the synchronization of spiking networks with conductance-based synapses, we used the general form for a conductance-based neuron with conductance-based coupling.  The network equations are given by: 
\begin{eqnarray}
C\frac{dV_i}{dt} &=& F(V,\bm x_i) - \sum_{j=1}^{N} g_{ij}r_j(t)(V_i - E) \label{sc1}\\
\frac{d\bm x_i}{dt} &=& G(V_i,\bm x_i) \label{e2} \\
\frac{dr_i}{dt} &=& a_r T(V_i)(1-r_i) - a_d r_i\label{sc2},
\end{eqnarray} 
where $V$ corresponds to the voltage of the neuron, and $\bm x$ consists of a vector of gating variables.   Although our derivation was for a general conductance-based neuron, we restricted our numerical analysis primarily to the Morris-Lecar neuron model \cite{ml1,bardtext} which we describe below.  These neurons are coupled with a smooth synaptic gating variable equation (\ref{sc2}) which was first introduced in \cite{destexhe}.  The synaptic connection between neuron $j$ to neuron $i$ is given by $\bar{g}_{ij}r_j(t)(V_i-E)$, where the inhibitory/excitatory nature of the synapse is determined by the reversal potential $E$.   The function $T(V_j)$ models the amount of neurotransmitter released in the synaptic cleft by neuron $j$, and is given by:
\begin{eqnarray*}
T(V) = \frac{\bar{T}_{max}}{1 + \exp\left(-(V- V_T)/K_p\right)}
\end{eqnarray*}

Finally, the conductance matrices $\bm g$ is an $N\times N$ matrix with the following constraints:
\begin{eqnarray}
g_{ij} \geq 0, \forall i,j =1,2,\ldots N, \quad \sum_{j=1}^N \bar{g}_{ij} = G, \forall i=1,2,\ldots N \label{const1} 
\end{eqnarray}
Thus, all the unitary conductances must be positive.  The constraints \ref{const1} are critical for the application of a modified Master Stability Function (MSF) analysis of the synchronized solutions.  

With the constraint (\ref{const1}) in hand, then the synchronous solution corresponds to the dynamics of a self-coupled or autaptic neuron:
\begin{eqnarray*}
C\frac{dV_S}{dt} &=& F(V_S,\bm x_S) - \bar{g}r_S(t)(V_S - E_E) \label{sse1} \\
\frac{d\bm x_S}{dt} &=& G(V_S,\bm x_S) \\
\frac{dr_S}{dt} &=& a_r T(V_S)(1-r_S) - a_d r_S \label{sse3} 
\end{eqnarray*} 
where $(V_S(t),\bm x_S(t),r_S(t))$ is the solution to the synchronous differential equations (\ref{sse1})-(\ref{sse3}).

Next, we perturbed around the synchronous solution with: 

 $$V_i = \epsilon^{V}_i + V_S(t), \bm x_i = \epsilon^{\bm x}_i + \bm x_S(t), r^E_i = \epsilon^{r^E}_i + r_S(t)$$ 

 which yielded the following linearization
\begin{eqnarray}
C\frac{d\bm \epsilon^{V}}{dt} &=& \left(\frac{\partial F}{\partial V}-\bar{g} r_S\right) \bm \epsilon^{V} + \sum_{j=1}^m  \frac{\partial F}{\partial x_j}\bm \epsilon^{x_j} -(V_S - E) \bm g \bm \epsilon^r \label{lp1} \\ 
\frac{d\bm \epsilon^{\bm x_i}}{dt} &=& \frac{\partial G}{\partial V}\bm \epsilon^V + \sum_{j=1}^m \frac{\partial G_i}{\partial x_j} \bm \epsilon^{x_j} \\ 
\frac{d\bm \epsilon^{r^E}}{dt} &=& a_r T'(V_s)(1-r_s)\bm \epsilon^{V} - (a_r T(V_S)+a_d)\bm \epsilon^r \label{lp3}
\end{eqnarray} 
Note that all of the partial derivative terms (e.g. $\frac{\partial F}{\partial V}, \frac{\partial G}{ \partial x_j}$) are evaluated with respect to the synchronized solution $(V_S(t),\bm x_S(t),r_S(t))$.    

Next, we will make the standard assumption in MSF-applications that the matrix $\bm g$ is diagonalizable 
\begin{eqnarray*}
\bm g = \bm P \bm D \bm P^{-1}
\end{eqnarray*}
Then consider the substitution:
\begin{eqnarray}
\bm \eta^V = \bm P^{-1} \bm \epsilon^V, \quad \bm \eta^{x_i} = \bm P^{-1} \bm \epsilon^{x_i}, \quad \bm \eta^r = \bm P^{-1} \bm \epsilon^r \label{sub}.
\end{eqnarray} 
Substituting (\ref{sub}) into equations (\ref{lp1})-(\ref{lp3}) yields:
\begin{eqnarray*}
C\frac{d\bm \eta^{V}}{dt} &=& \left(\frac{\partial F}{\partial V}-\bar{g} r_S\right) \bm \eta^{V} + \sum_{j=1}^m  \frac{\partial F}{\partial x_j}\bm \eta^{x_j} - (V_S - E) \bm D\bm  \eta^r \\ 
\frac{d\bm \eta^{\bm x_i}}{dt} &=& \frac{\partial G}{\partial V}\bm \eta^V + \sum_{j=1}^m \frac{\partial G_i}{\partial x_j} \bm \eta^{x_j} \\ 
\frac{d\bm \eta^{r^E}}{dt} &=& T'(V_S)(1-r_S)\bm \eta^{V} - (a_r T(V_S)+a_d)\bm \eta^r .
\end{eqnarray*} 
The key insight drawn from the MSF approach is that equation (\ref{dp1}) is now effectively uncoupled as the system has been block diagonalized (see Appendix 1 for further details).  This implies that to determine the stability of the synchronized solution, we can determine Lyapunov exponents of the system 
\begin{eqnarray}
 C\frac{d\eta^{V}}{dt} &=& \left(\frac{\partial F}{\partial V}-\bar{g} r_S\right)  \eta^{V} + \sum_{j=1}^m  \frac{\partial F}{\partial x_j}\eta^{x_j} - \lambda_i(V_S-E) \eta^r \label{dp1} \\ 
\frac{d\eta^{x_i}}{dt} &=& \frac{\partial G}{\partial V} \eta^V + \sum_{j=1}^m \frac{\partial G_i}{\partial x_j} \eta^{x_j} \\ 
\frac{d \eta^{r}}{dt} &=& a_rT'(V_S)(1-r_S)\eta^{V} - (a_r T(V_S)+a_d) \eta^r \label{dp3}.
\end{eqnarray}
over a mesh in $\lambda$.  The system (\ref{dp1})-(\ref{dp3}) is numerically integrated along with the synchronous solution (\ref{sse1})-(\ref{sse3}), in conjunction with a numerical estimation of the Lyapunov exponents (see Supplementary Information for Details).   Then, the stability the synchronized solution can be determined readily by first computing eigenvalues of the connectivity matrix, and then ``looking up" the values on the mesh produced by the MSF 
.  

\subsection{\textbf{The Morris-Lecar Neuron Model}}

To test the predictions of the MSF function, we primarily considered the Morris-Lecar neuron model \cite{ml1,bardtext} : 
\begin{eqnarray*}
C \frac{dV}{dt} &=& I - g_L (V - E_L ) - g_Kn(V-E_K)  -g_{Ca} m_\infty(V)(V-E_{Ca}) \\
\frac{dn}{dt}&=& \phi \left(\frac{n_{\infty}(V) - n}{\tau_n(V)} \right)\\ 
m_\infty(V)&=& \frac{1}{2}(1+\tanh(V-V_1)/V_2)\\
n_\infty(V)&=&\frac{1}{2}(1+\tanh(V-V_3)/V_4)\\
\tau_n(V) &=& \frac{1}{\cosh(V-V_3)/(2V_4)}
\end{eqnarray*}

The parameters for this model are given in Table \ref{table1} and correspond to two classical regimes, the Hodgkin Class I regime which corresponds to a Saddle Node on an Invariance Cycle (SNIC) bifurcation from quiescence to spiking, and the Hodgkin Class II regime which corresponds to a subcritical Hopf bifurcation followed by a saddle-node of limit cycles from quiescence to spiking \cite{bardtext}.  The network equations are given by 

\begin{eqnarray*}
C \frac{dV_i}{dt} &=& I - g_L (V_i - E_L ) - g_Kn_i(V_i-E_K)  -g_{Ca} m_\infty(V_i)(V_i-E_{Ca})-\sum_{j=1}^N g_{ij} r_j(t)(V_i - E) \\
\frac{dn_i}{dt}&=& \phi \left(\frac{n_{\infty}(V_i) - n_i}{\tau_n(V_i)} \right)\\ 
\frac{dr_i}{dt} &=& a_rT(V_i)(1-r_i) - r_i a_d \\
\sum_{j=1}^N g_{ij} &=& \bar{g},\forall i, \quad g_{ij}\geq0  
\end{eqnarray*}
while the synchronized solution corresponds to the autaptic Morris-Lecar neuron:
\begin{eqnarray}
C \frac{dV_S}{dt} &=& I - g_L (V_S - E_L ) - g_Kn_S(V_S-E_K)  -g_{Ca} m_\infty(V_S)(V_S-E_{Ca})-\bar{g} r_S(t)(V_S - E) \label{ml1}\\
\frac{dn_S}{dt}&=& \phi \left(\frac{n_{\infty}(V_S) - n_S}{\tau_n(V_S)} \right)\\ 
\frac{dr_S}{dt} &=& a_rT(V_S)(1-r_S) - r_S a_d \label{ml3} 
\end{eqnarray}

\begin{figure}[htp!] 
\centering
\includegraphics[scale=0.5]{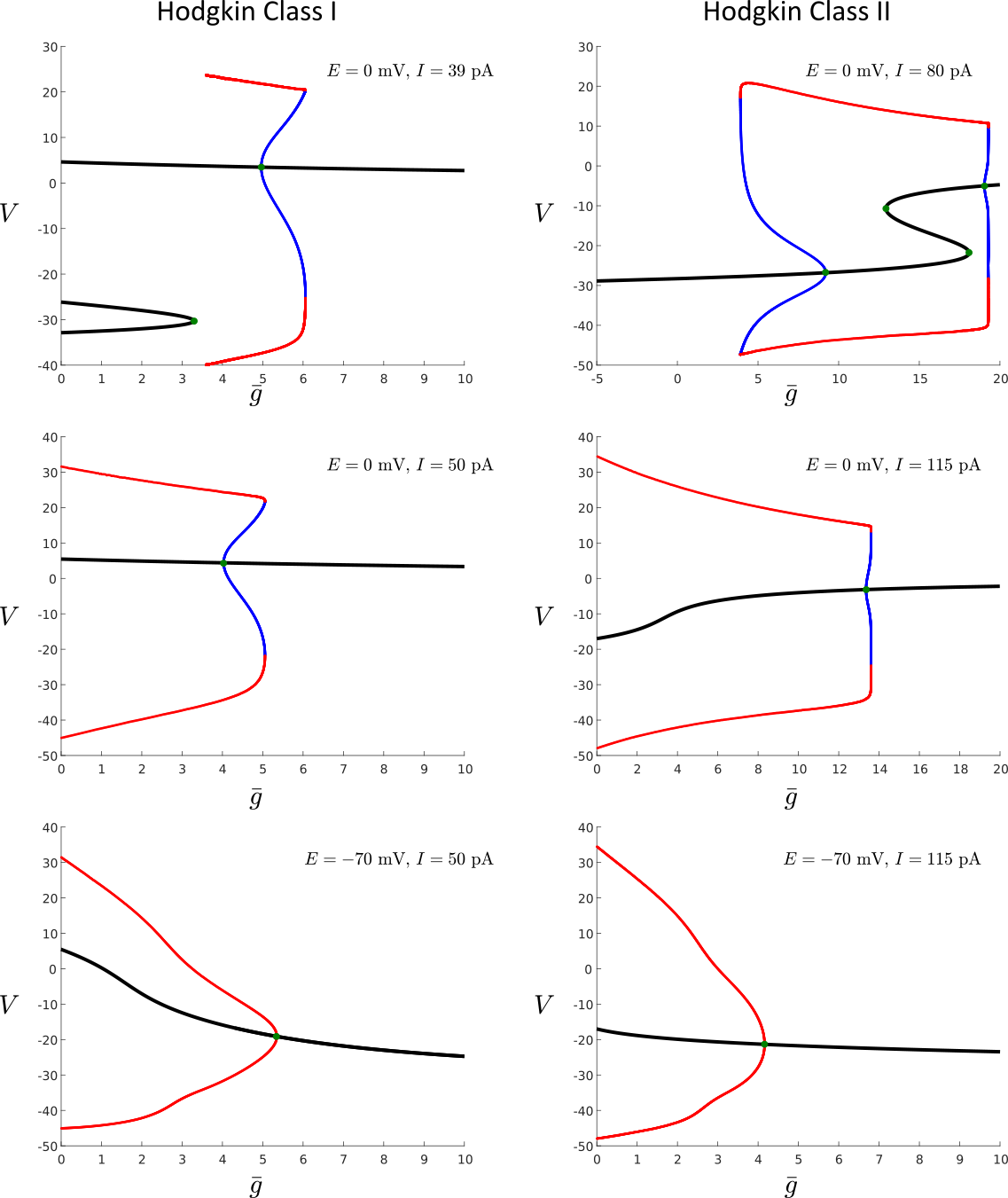}
\caption{The bifurcation diagrams for self-coupled Morris-Lecar neuron models.  The left column corresponds to the Hodgkin Class I parameter regime, while the right column corresponds to the Hodgkin Class II parameter regime.  The parameters correspond to excitatory self-coupling near the onset to spiking (low $I$), excitatory self-coupling far from the onset to spiking (high $I$), and inhibitory self-coupling far from the onset to spiking (high $I$).  The inhibitory self-coupling regime near the onset to spiking was not considered as it typically leads to narrower parameter regions in $\bar{g}$, or can sometimes lead to spiking in non-physical regimes (e.g. $\bar{g}<0$).  The master stability functions were computed for Hodgkin Class I and Hodgkin Class II parameters in the large $I$ regimes for similar ranges in $\bar{g}$. }\label{f1}
\end{figure}

Prior to computing the MSF, we investigated which regions of $\bar{g}$ lead to stable spiking solutions as a function of the driving current $I$, and the reversal of the synapses $E$ for the synchronous solution in equations (\ref{ml1})-(\ref{ml3}).  First, we found that excitatory self-coupling ($E= 0$ mV) did not change the overall bifurcation types for Hodgkin Class I or Hodgkin Class II parameter regimes (Figure \ref{f1}).  However, we did find changes to the overall bifurcation structure with inhibitory self-coupling would either shift the spiking regimes to non-physical negative $\bar{g}$, or reduce these regimes to narrow parameter ranges in $\bar{g}$ depending on how close the driving current was to the bifurcation point of the non-autaptic neuron.  Thus, we primarily focused on parameter regimes with higher driving currents for both excitatory ($E=0$ mV) and inhibitory ($E=-70$ mV) reversal potentials.   Furthermore, we found that for these regimes, we could systematically vary $\bar{g}$ over the interval $(0,3]$ across both Class I and Class II parameter regimes, albeit with different applied currents ($I=50$ pA for Class I, $I = 115$ pA for Class II).  For the currents we considered, the local bifurcation structure appears largely identical for both classes of firing (Figure \ref{f1}), with both parameter regimes stopping spiking via a supercritical Hopf ($E=-70$ mV) bifurcation for $\bar{g}>3$ or subcritical Hopf ($E=0$ mV) regimes. 

\subsection{\textbf{Computing and Validating the Master Stability Function for the Morris-Lecar Model}} 
The block linearization for the MSF function of the Morris-Lecar network is given by 
\begin{eqnarray}
C\frac{d\eta^V}{dt} &=& \left(-g_L - g_{Ca}m^{\infty}(V_S) - g_{Ca}(V_S-E_{Ca}) \frac{dm^{\infty}}{dV} g_K n_S -\bar{g} r_S\right)\eta_V\nonumber\\& -& g_K(V_s- E_k)\eta^n - \bar{g}\lambda(V_S-E)\eta^r \label{lameq1}\\
\frac{d\eta^n}{dt} &=& \phi\left(\frac{1}{\tau_n(V_S)}\frac{dn^\infty(V_S)}{dV} -\frac{d\tau_n}{dV_S}\frac{n_\infty(V_S) - n_S}{\tau_n(V_S)^2}\right)\eta_V - \frac{\phi}{\tau_n(V_S)}\eta^n \\
\frac{d\eta^r}{dt}&=& a_r \frac{dT(V_S)}{dV}(1-r_s)\eta^V - (a_r(T(V_S)) + a_d)\eta^r \label{lameq2}
\end{eqnarray} 
where $(V_S,\eta_S,r_S)$ correspond to the synchronized solution for the self-coupled node:

To compute the master stability function, we simulated the system of equations (\ref{ml1})-(\ref{ml3}) while computing the Lyapunov exponents in parallel for each of the blocks in (\ref{lameq1})-(\ref{lameq2}).  The eigenvalues $\lambda$ were selected over a $101\times101$ mesh over the unit cube $[a,b] \in [0,1]^2$ with $\lambda = a+bi$.  

\begin{figure}[htp!] 
\centering
\includegraphics[scale=0.75]{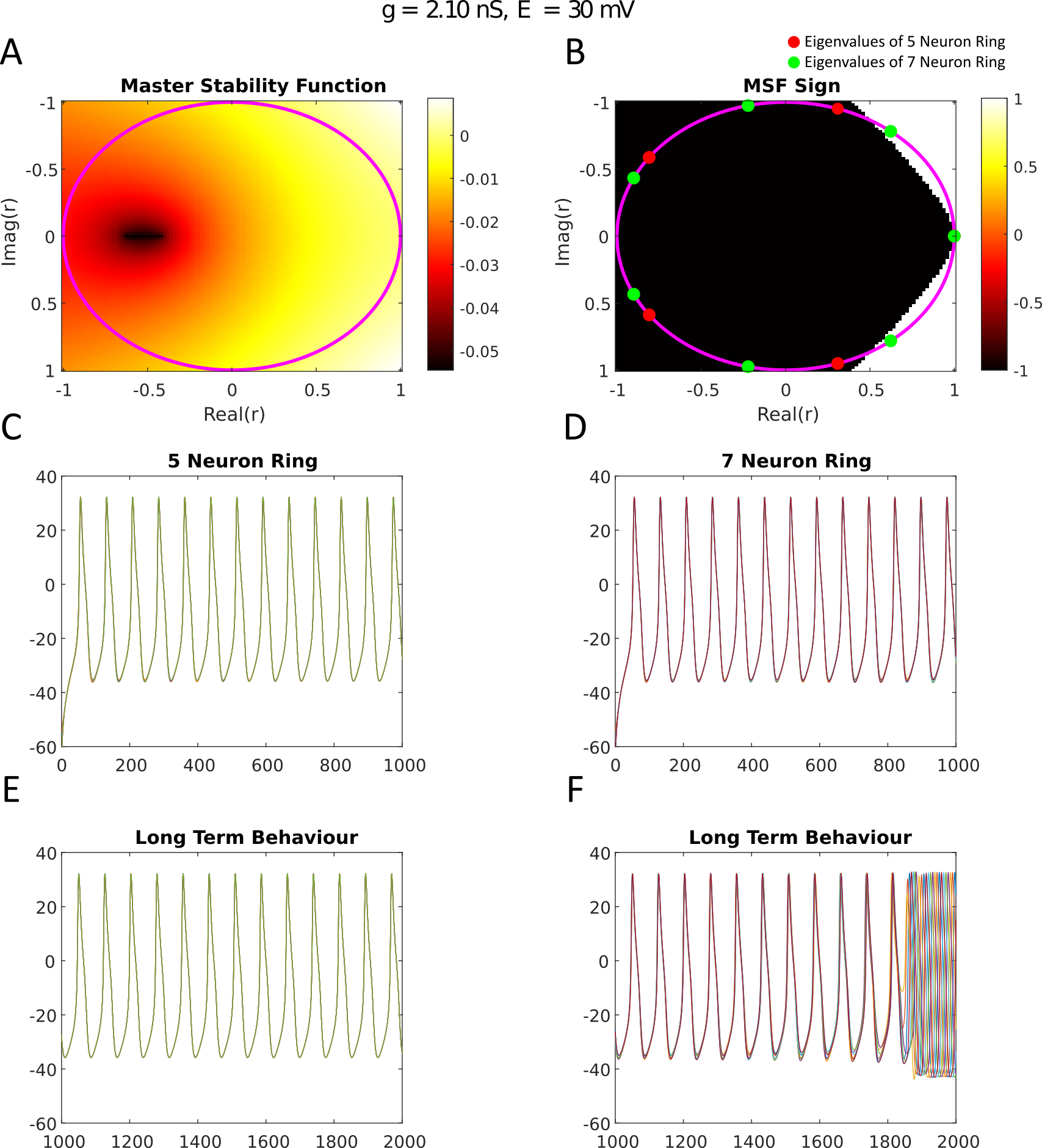}
\caption{Testing the master stability function in simulated ring networks \textbf{(A)} The MSF for the Morris-Lecar neuron under Hodgkin Class I Excitability.  The MSF was computed over a $101\times 101$ evenly distributed mesh over the complex plane for $\bar{g}=2.1$ nS and $E = 30$ mV (excitatory synapses) \textbf{(B)} The sign of the MSF function.  Black denotes negative values, indicating a stable Lyapunov exponent associated with the eigenvalue, while white denotes positive Lyapunov exponents associated with the eigenvalue.  The MSF was tested with two rings, a 5-neuron ring and a 7 neuron ring.  The 6-neuron ring lay close to the sign change transition point.   \textbf{(C)} The 5-neurons in the ring are simulated near the basin of attraction of the synchronous solution.  The voltage of the neurons is initialized with a normally distributed random variable with mean -60 mV, and a standard deviation of 5 mV.  The $n$ and $r$ variables are all set to 0.5.   \textbf{(D)} Identical as in (A), only with the 7 neuron ring.  \textbf{(E)} The asymptotic behaviour of the 5-neuron ring is a synchronous solution.   \textbf{(F)} The 7 neuron ring desynchronizes after a suitably long period of time.  }
\label{f2}
\end{figure}

A single computation for the MSF  for $\bar{g}=2.1$ nS and $E = 30$ mV is shown in Figure \ref{f2}.   For these particular parameter regimes, the reversal potential indicates a predominantly excitatory synaptic coupling.   This excitatory connectivity leads to large regions in the eigenvalue space where the MSF is negative, (Figure \ref{f2}A-B).   For eigenvalues with larger magnitudes, and positive real components, the MSF exhibits a sign change indicating a loss of stability in the synchronous solution.  To test this loss of stability, we used ring networks of different sizes as the eigenvalues of a ring of $N$ neurons lie on the unit circle as the $N$th roots of unity. A ring with 5-neurons has all eigenvalues laying in the negative MSF area indicating all negative Lyapunov exponents while a ring with 7 neurons has a pair of complex conjugate eigenvalues in the positive MSF area.  This indicates that the synchronous solution is stable for a ring of $N=5$ neurons but unstable for $N=7$ neurons.  The ring of $N=6$ neurons has eigenvalues that are extremely close to the sign change of the MSF, and thus was not considered.  The $N=5$ Morris-Lecar ring and the $N=7$ Morris-Lecar ring were both simulated with initial conditions near the synchronous solution to test its local asymptotic stability (Figure \ref{f2}C-D).   After the initial transient, the $N=5$ ring converges to the synchronous solution (Figure \ref{f2}E) while the $N=7$ ring diverges (Figure \ref{f2}F).

\subsection{\textbf{The Master Stability Function over the ($E,\bar{g}$) parameter space for Class I and Class II Neurons}} 

\begin{figure}[htp!] 
\centering
\includegraphics[scale=0.45]{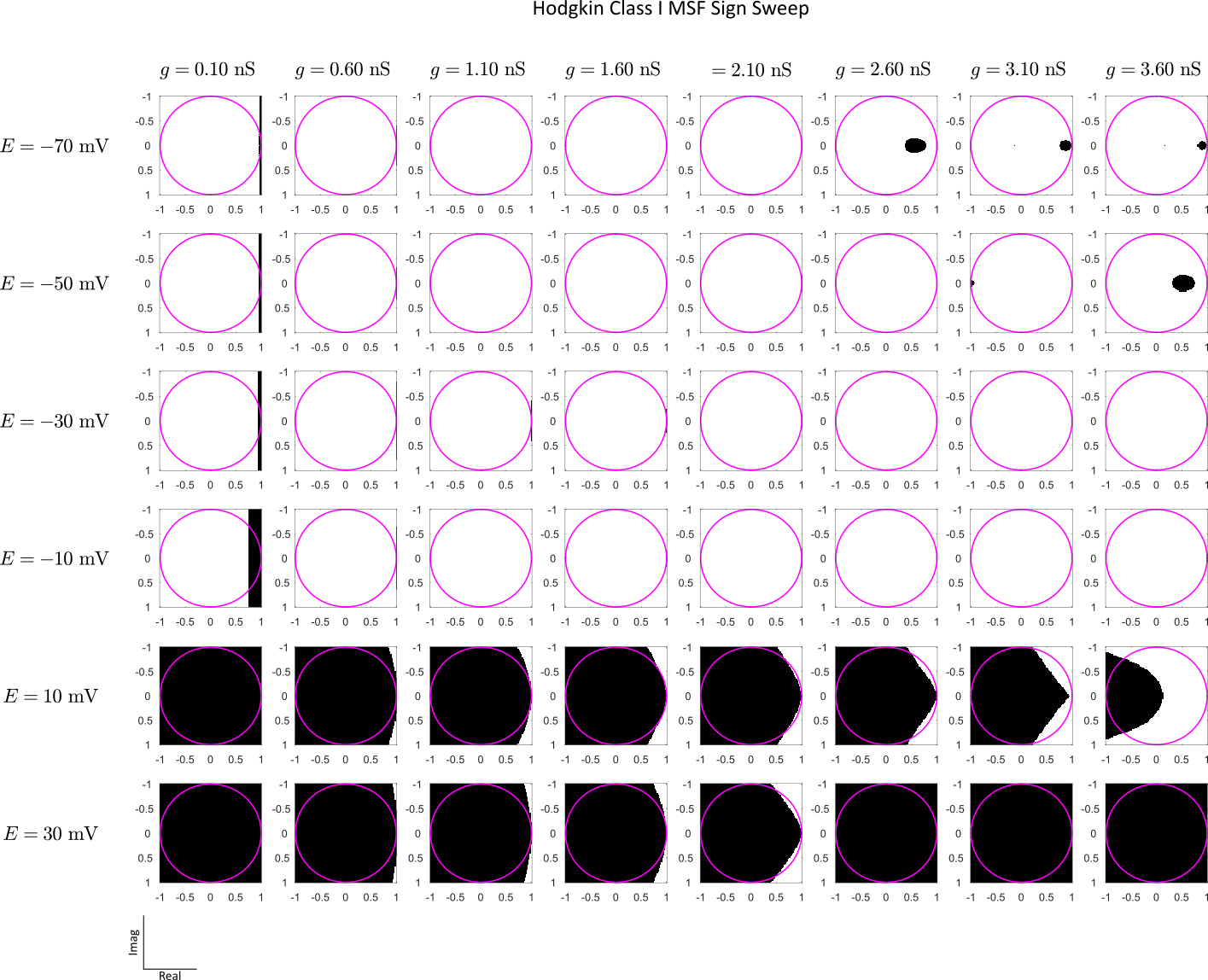}
\caption{The sign of the master stability function (MSF) as the network parameters $\bar{g},E$ are varied  The Morris-Lecar parameters were taken to be in the Hodgkin Class I parameter regime.    The sign of the MSF function was computed over a mesh in the $(\bar{g},E)$ parameter space.  The mesh points correspond to $E = -70 + 20j$ mV, for $j = 0,1,2,\ldots 5$ and$g = 0.1 + 0.5k$ nS for $k=0,1,2\ldots 7$.  In between $E=-30$ and $E=-10$ mV, there is a large-scale transition from predominantly unstable synchronized solutions and predominantly stable synchronized solutions.   }\label{f3}
\end{figure}

\begin{figure}[htp!] 
\centering
\includegraphics[scale=0.45]{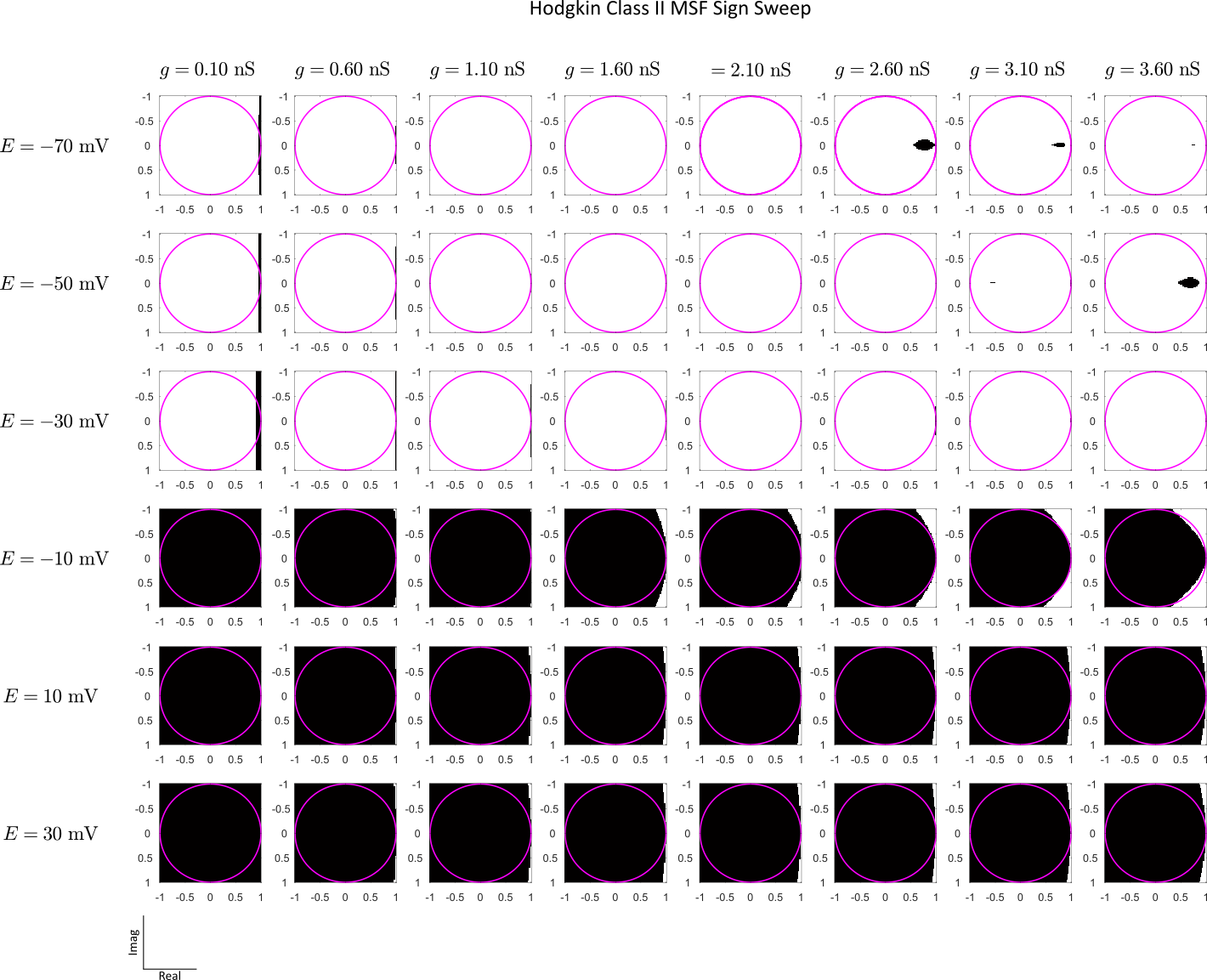}
\caption{The sign of the master stability function (MSF) as the network parameters $\bar{g},E$ are varied.  The Morris-Lecar parameters were taken to be in the Hodgkin Class II parameter regime.  The sign of the MSF function was computed over a mesh in the $(\bar{g},E)$ parameter space.  The mesh points correspond to $E = -70 + 20j$ mV, for $j = 0,1,2,\ldots 5$ and$g = 0.1 + 0.5k$ nS for $k=0,1,2\ldots 7$.  In between $E=-10$ and $E=10$ mV, there is a large-scale transition from predominantly unstable synchronized solutions and predominantly stable synchronized solutions.   Note the similarities to Figure \ref{f3} }\label{f4}
\end{figure}

Next, we determined how the inhibitory/excitatory valence of the conductance-based synapse would impact synchronizability.  For a sufficiently large positive reversal potential, the current induced by presynaptic spikes primarily serves to depolarize the cell, and therefore initiates subsequent spikes.  For a sufficiently large negative reversal potential, the cell becomes hyperpolarized by presynaptic spikes.  
  The conventional wisdom would be that increasing the reversal potential would lead to more synchronization as all the synapses transitioned from inhibitory to excitatory.  To test this hypothesis, we computed the MSF over a discrete mesh in $(\bar{g},E)$ space for the Morris-Lecar network under both Class I (Figure \ref{f3}) and Class II parameters (Figure \ref{f4}).  

We found that the conventional wisdom largely prevails here, with increasing reversal potentials leading to synchronizability (Figure \ref{f3}-\ref{f4}).  However, there are some important caveats and unexpected deviations that occur. First, we noticed that predominantly inhibitory connectivity ($E\leq -50 mV$) can lead to islands of synchronizability when the global conductance strength $\bar{g}$ is sufficiently high. This is a common feature of MSFs where one can find localized and compact region(s) of stability \cite{msf3}.  These islands are ``born" and die through a process we describe below.  Second, the transition from a predominantly inhibitory synapse to an excitatory synapse appears quite suddenly, somewhere in between $E\in [-10,10]$ mV.  For the most part, there does not appear to be a gradual change in synchronizability, but an abrupt one.  Finally, even for ``excitatory" synapses, increasing the global conductance strength can decrease the area where the MSF is negative.  This occurs for example in the $E=10$ mV reversal potential, where a higher $\bar{g}$ progressively erodes the synchronizability regime.  Interestingly, the loss of stability with increasing $\bar{g}$ can also be reversed for a sufficiently high reversal potential.   Finally, we note that in the weak coupling regime ($\bar{g}\ll O(1)$), there is a vertical slice of synchronizability where neurons can synchronize provided that they connect primarily with strong autaptic connections and weak cross connections to each other

\begin{figure}[htp!] 
\centering
\includegraphics[scale=0.8]{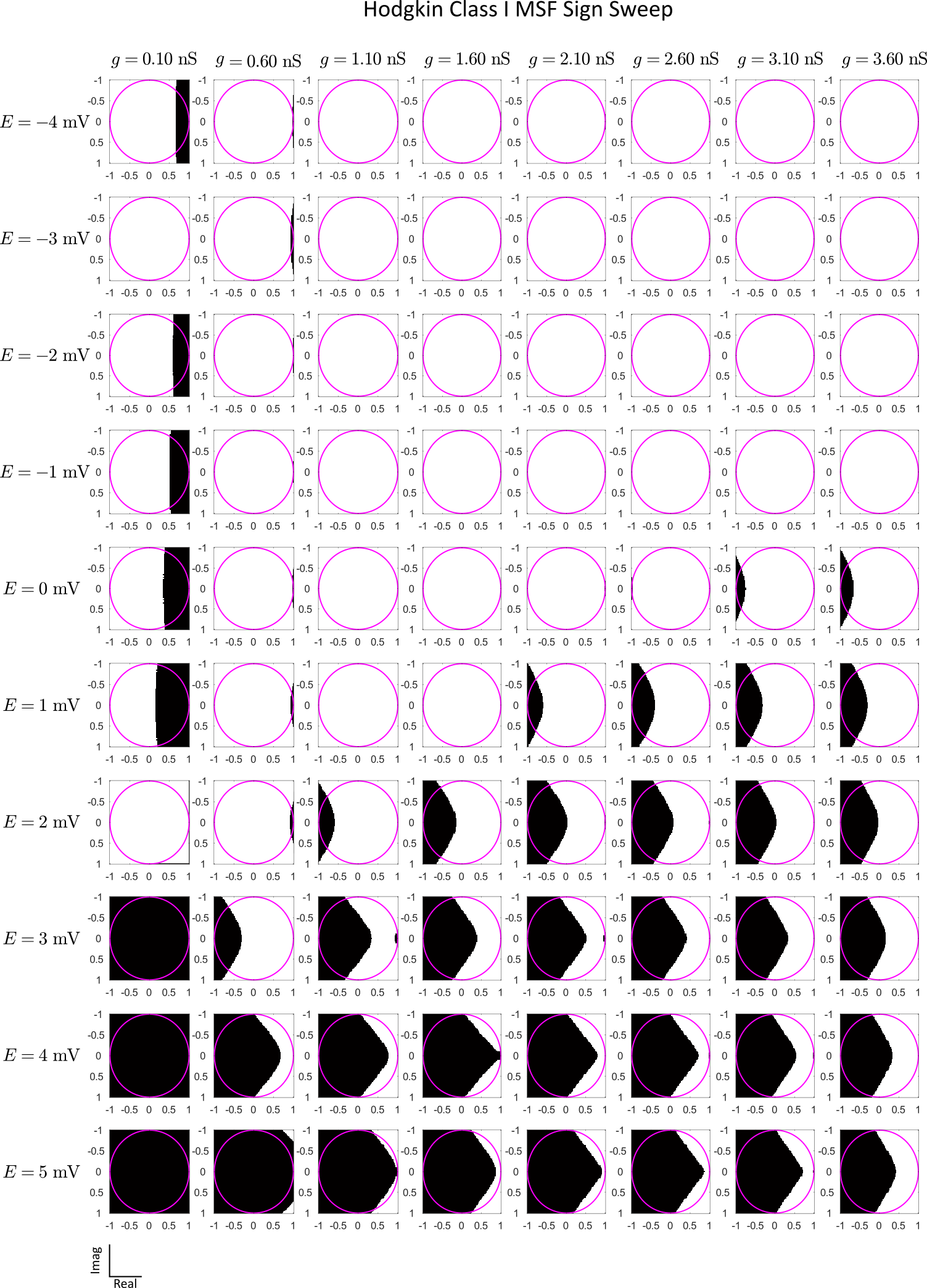}
\caption{The sign of the master stability function (MSF) as the network parameters $\bar{g},E$ are varied.  The Morris-Lecar parameters were taken to be in the Hodgkin Class II parameter regime.  The sign of the MSF function was computed over a mesh in the $(\bar{g},E)$ parameter space.  The mesh points correspond to $E = -4 + j$ mV, for $j = 0,1,2,\ldots 9$ and$g = 0.1 + 0.5k$ nS for $k=0,1,2\ldots 7$.  In between $E=-4$ and $E=5$ mV, there is a large-scale transition from predominantly unstable synchronize solutions and predominantly stable synchronized solutions.   }\label{f5}
\end{figure}

\subsection{\textbf{Abrupt Changes to the MSF in the ($E,\bar{g}$) parameter space.}}

For both Hodgkin Class I and Hodgkin Class II parameter sets, there was a near global change in the sign of the MSF.  For Class I excitability, the MSF evolved from predominantly positive Lyapunov exponents for $E=-10$ mV, to predominantly negative Lyapunov exponents for $E=10$ mV.  We investigated how abrupt this transition was by considering the Hodgkin Class I parameters on a finer mesh.  We computed the MSF and plotted its sign from the interval $E=-4$ mV to $E = 5$ mV with increments of 1 mV, with $\bar{g}$ fixed as in Figures \ref{f3}-\ref{f4} (Figure \ref{f5}).  We found that the MSF was predominantly positive, aside from the small $\bar{g}$ region.  As $E$ is increased however from $E=0$ to $E=5$ mV the MSF rapidly changes sign over a large area of the admissible eigenvalue space $|\lambda | \leq 1$.   For every millivolt of increase in the reversal potential, the region of stability for the synchronous solution grows rapidly from the left side of the admissible eigenvalue domain. For non-weak coupling $(\bar{g}\gg O(10^{-1}) nS$, most admissible weight matrices go from unstable synchronous solutions to stable synchronous solutions with the change of just a few millivolt's in the reversal potential of the synapse.

\begin{figure}[htp!] 
\centering
\includegraphics[scale=0.35]{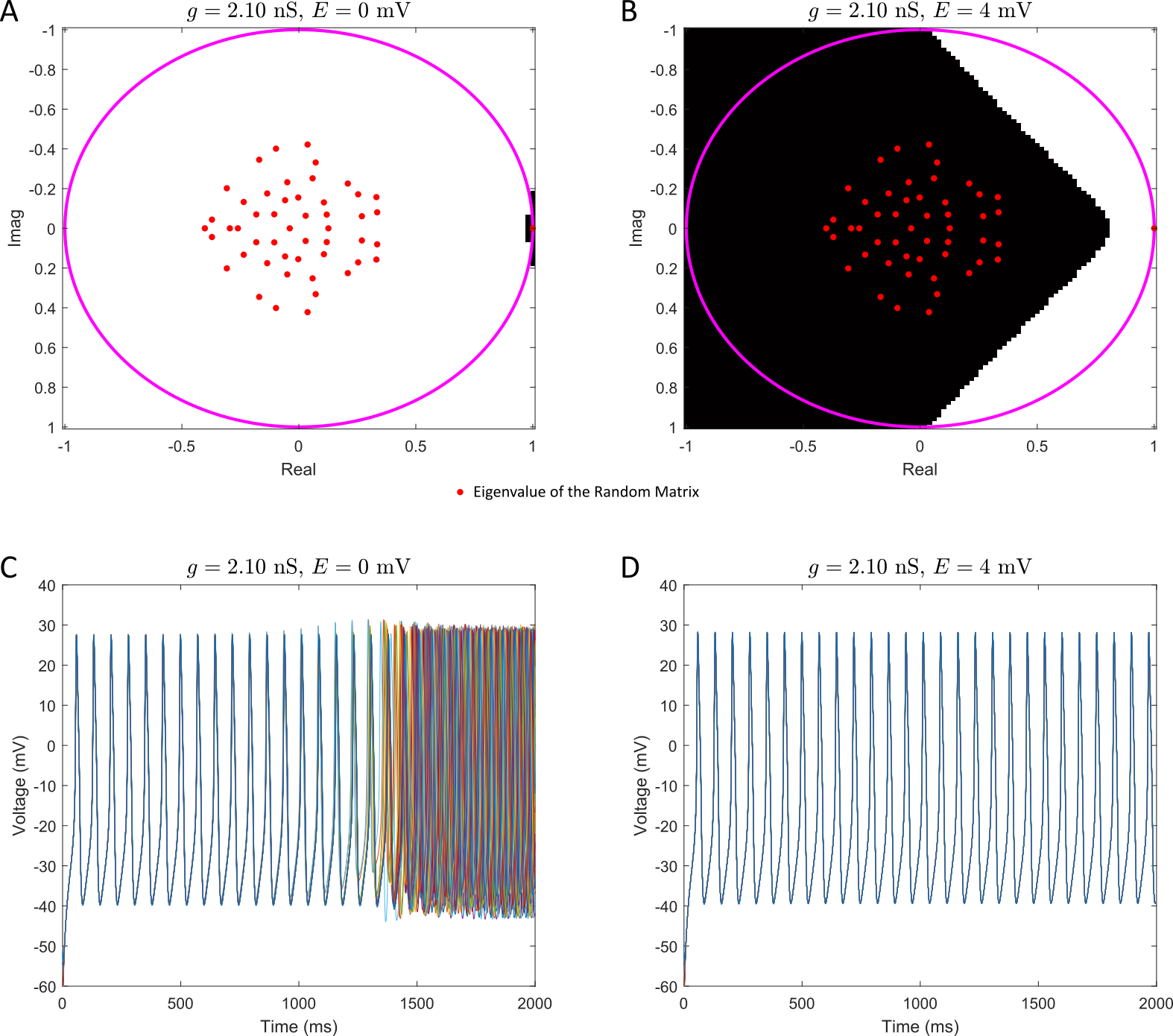}
\caption{A change of a few mV in the reversal potential changes the stability of the synchronous solution.  \textbf{(A)} The sign of the MSF function for $\bar{g}=2.1$, $E = 0$ mV.  The eigenvalues (red dots) of a randomly generated $N=50$ node Erdős–Rényi model network that is 85\% sparse is generated are also plotted.  \textbf{(B)} The sign of the MSF function for $\bar{g}=2.1$, $E = 4$ mV.  The eigenvalues (red dots) are identical as in (A).  \textbf{(C)} A simulation of a network Morris-Lecar neurons with $\bar{g}$ and $E$ identical as in (A), coupled with the weight matrix from (A).  Every neuron is generated with $n_j = 0.5$, $r_j = 0.5$, and a random initial voltage drawn from a normal distribution with a mean of -60 mV, and standard deviation of 3 mV. \textbf{(D)}  An identical simulation as in (C), only with $\bar{g}=2.1$ and $E=4$ mV.  The initial conditions are identical as in (C).  The synchronous solution is now locally asymptotically stable.   The network parameters were in the Hodgkin Class I regime. }\label{f6}
\end{figure}

Next, we tested these abrupt changes in synchronizability by using randomly coupled networks of Morris-Lecar neurons (Figure \ref{f6}).    The network was simulated at two parameter values, with $\bar{g} =2.1$ nS and $E = 0$ mV, and $\bar{g}=2.1$ nS and $E=4$ mV.  The former was predicted to yield unstable synchronized solutions while the latter had a large region of stability.  The network consisted of 50 neurons, with 85\% sparse random coupling, and all gating variables initialized to 0.5, while the voltage variable was drawn from a normal random variable with a mean of -60 mV and a standard deviation of 3 mV (Figure \ref{f6}A-B).  The eigenvalues of the matrix were verified to lie in the positive sign MSF ($E=0$ mV) and the negative sign MSF ($E=4$ mV), respectively  (Figure \ref{f6}A-B).  The network simulated with the lower reversal potential desynchronized after ~1.5 seconds while the network at the slightly higher reversal potential synchronized (Figure \ref{f6}C-D).   These results demonstrate how a change of only a few millivolts in the reversal potential changes the stability of the synchronous solution for many types of network coupling.

\subsection{\textbf{Islands of Synchronizability in the ($E,\bar{g}$) parameter space.}}

\begin{figure}[htp!] 
\centering
\includegraphics[scale=0.32]{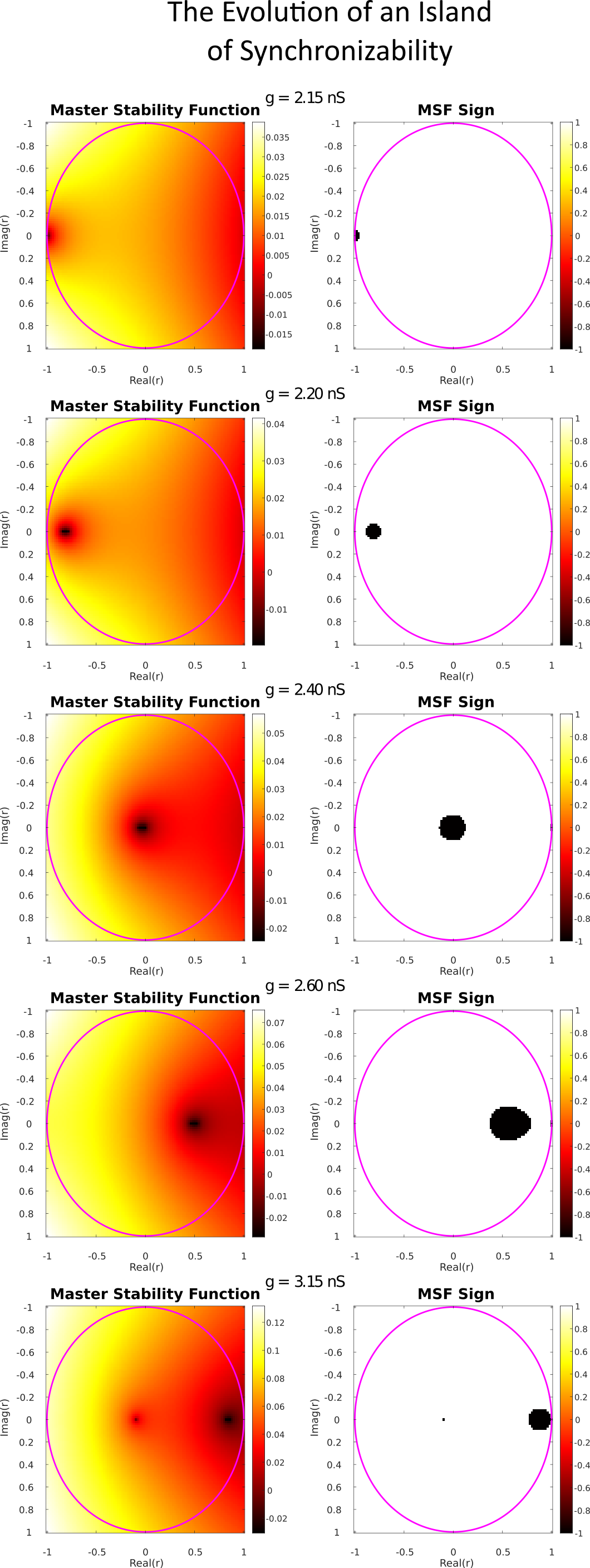}
\caption{The ``birth" and ``death" of an island of synchronizability as $\bar{g}$ is increased.  The MSF was computed for the Morris-Lecar neuron model under Hodgkin Class I excitability.  The island of stability emerges from the negative real part of the eigenvalue mesh (1st row).  Then, the island undergoes a period of expansion and lateral movement to larger real components (2nd to 4th rows).  The island subsequently collides with neutral stability eigenvalue $\lambda = 1$ and begins to be  ``absorbed" by it (5th row).  A second, very small island of synchronizability has emerged in the last column.  The reversal potential was -70 mV. }\label{f7}
\end{figure}

To witness the evolution of an island of synchronizability, we first computed the MSF function over a finer mesh in the $\bar{g}$ space with $E$ fixed to $-70$ mV (Figure \ref{f7}).  In these islands, the MSF has a local minimum which can emerge from the left-hand side of the eigenvalue domain (Figure \ref{f7}).  The island transitions to the right-hand side of the eigenvalue domain while expanding in size.  Eventually, the island is ``absorbed" by the neutral stability eigenvalue at $\lambda = 1$ (Figure \ref{f7}).   Multiple islands (negative minima of the MSF function) can co-exist (Figure \ref{f7}). 

\begin{figure}[htp!] 
\centering
\includegraphics[scale=0.35]{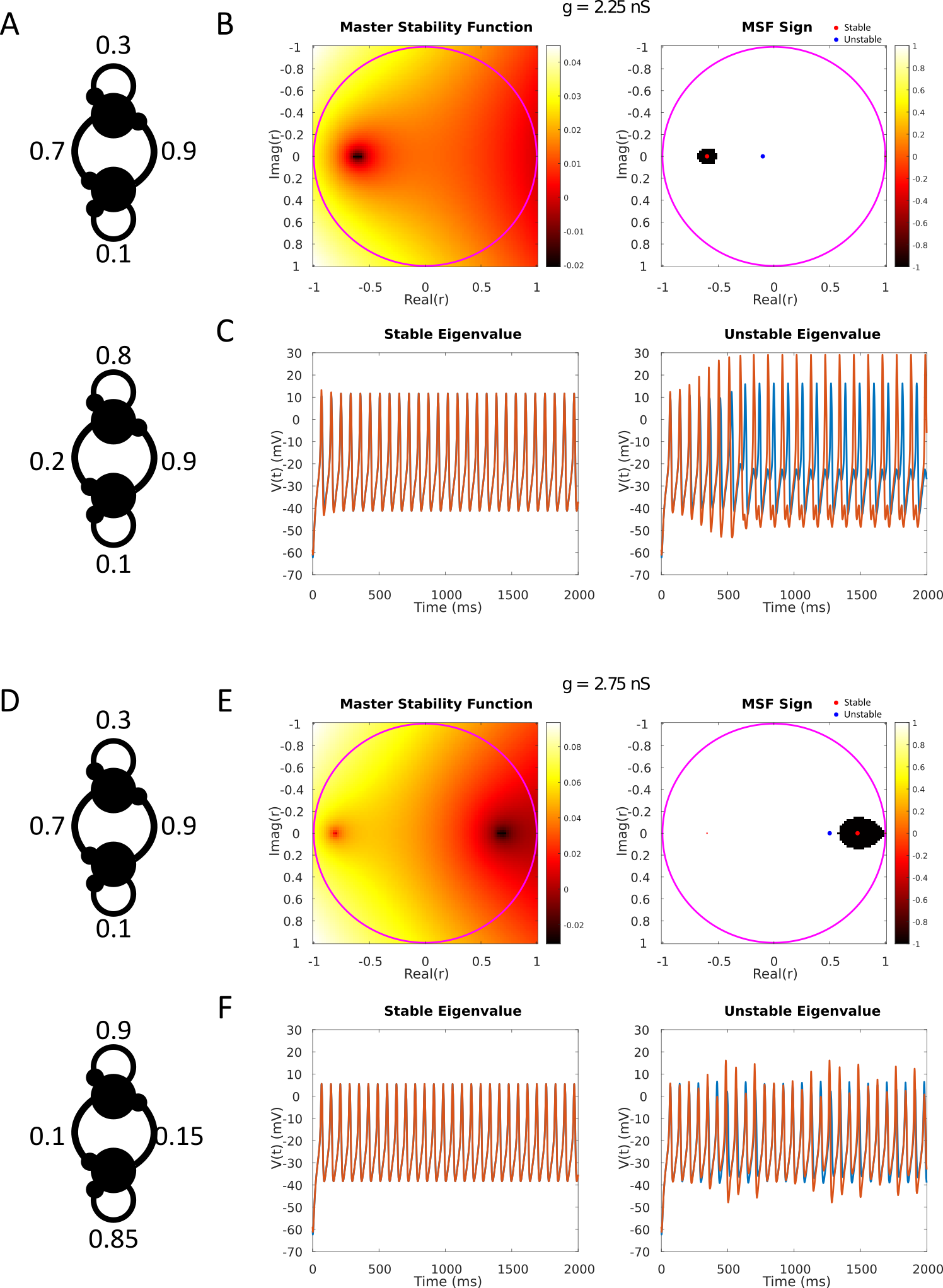}
\caption{Testing the islands of stability with simulated pairs of coupled oscillators. \textbf{(A)} Two connectivity matrices for a network of two coupled Morris-Lecar neurons with Hodgkin Class I excitability.  The two matrices yield eigenvalues that lie in the island of synchronizability (top), or are outside and adjacent to it (bottom).  Note that these matrices satisfy both the unity row-sum constraint, and all positive elements constraint of a conductance matrix.    \textbf{(B)} The computed master stability function for $\bar{g}=2.25$ nS (left), $E=-70$ mV, and the sign of the MSF (right).  Both matrices have eigenvalues of 1 (neutral stability) and a second eigenvalue less than 1.  The matrix on the top of (A) has a second eigenvalue within the island of stability (red), while the matrix on the bottom has an eigenvalue outside of the island of stability (blue). \textbf{(C)}  Simulation of the Morris-Lecar neurons coupled as in (A), with only the voltage plotted.  The matrix with an eigenvalue inside of the island has a locally asymptotically stable synchronous solution. The matrix on the bottom of (A) has an eigenvalue outside of the island, and has an unstable synchronous solution.   \textbf{(D)}  Identical as in (A), only with a larger conductance value $\bar{g}=2.75$ nAS. \textbf{(E)} Identical as in (B), with the connectivity matrices determined by (D) \textbf{(F)} Identical as in (C), with the connectivity matrices determined by (D). }\label{f8}
\end{figure}

Next, we investigated if these islands of synchronizability could be directly tested with a network simulation (Figure \ref{f8}).   As the islands bound small areas in the complex plane, we tested if networks of pairs $N=2$ of oscillators would synchronize (Figure \ref{f8}2A, 2D).  We constrained the connection strengths of these matrices to have a constant row sum of 1, which forces the maximum eigenvalue of $\lambda = 1$.  It is then trivial to constrain the connections such that the second eigenvalue lies in an arbitrary position on the real line (Figure \ref{f8} 2B, E).  For each parameter regime considered, two networks were tested.  One network was selected with an eigenvalue inside the island, and another with an eigenvalue outside of the island.  As predicted by the MSF function, we found that the connectivity matrices with eigenvalues within the islands of synchronizability lead to local asymptotic stability of the synchronous solution.   Given the small size of these islands, our results suggest that for conductance-based synapses, the connectivity graphs in which inhibition can induce synchronization may be highly constrained and strongly parameter dependent.

\section{Discussion} 

We have found that the synchronizability of conductance-based neurons is a complex affair.   By applying a modified MSF approach, we analyzed how networks of chemically coupled Morris-Lecar neurons synchronize in Class I and Class II regimes.  We found remarkable consistency in the stability regimes across these two parameter regimes.  As a general rule of thumb, higher reversal potentials tend to lead to the (local) stability of synchronous autaptic  solutions, and lower reversal potentials tend to lead to the (local) instability of synchronous autaptic solutions.  However, this rule of thumb is often deviated from as islands of stability under inhibition, and wedge-shaped regions of instability under excitation were both observed.   The actual change between predominantly stable or predominantly unstable synchronized solutions occurs very rapidly as a function of the reversal potential (a few mVs).   Evidently, the stability of synchronous solutions is strongly parameter dependent with small changes to the global conductance or small changes to the reversal potential stabilizing or destabilizing the synchronizability of different connectivity graphs. 

We note that we are not the only authors to consider applying the MSF function to neurons with  chemical synapses \cite{cjgj1,cjgj2}.  In \cite{cjgj1} the authors consider networks of Hindmarsh-Rose neuron models with both electrical and chemical coupling, thereby making a direct comparison to the work here difficult.   In \cite{cjgj2}, the authors consider networks of Izhikevich neuron models with chemical synapses and utilize non-smooth analysis to determine the MSF.  This was performed for electrical only, chemical only, and simultaneous electrical/chemical synapses. For chemical synapses, the estimated MSF appears positive  (Figure 5 in \cite{cjgj2}).  The modification to the MSF to allow for non-diffusive coupling appears in multiple sources in the literature \cite{cjgj2,w1,w2,w3,w4}. 

The islands of stability observed here occurred for primarily inhibitory (low $E$) connection strengths.  Interestingly, work with classical integrate-and-fire neurons with current based synapses has also demonstrated that inhibition can induce synchronization \cite{inhibs1}.   In particular, the authors consider leaky integrate-and-fire neurons with alpha-function like synaptic connectivity, where every spike at time $t^*$ increases the current arriving to a neuron by $E_s(t) = E_s(t-t^*)$ where $E_s = g\alpha^2 t \exp(-\alpha t)$.    The authors find that for inhibitory synapses ($g<0$), the synchronous solution is always stable, although at a critical value of $\alpha$ (faster synapses), a pitchfork bifurcation occurs that stabilizes for the asynchronous solution.  As the synapses become faster, the basin of attraction for the synchronous solution shrinks.  We primarily considered synapses with a rise time of $~1$ ($a_r = 1.1$)  ms and a decay time of $~5$ ms ($a_d = 0.19$).  For conductance-based synapses, it appears that the synchronous solution is stable under inhibition dominated regimes only for very specific connectivity graphs, in contrast with the behaviour of an inhibitory coupled integrate-and-fire network.  We leave the general analysis of the impacts of synaptic timing for future work, however we do remark that the findings in \cite{inhibs1} were supported and extended by subsequent modelling work  \cite{inhibs4} and even experimental findings \cite{inhibs2,inhibs3}.

By thoroughly exploring parameter space, we were able to shed insight into the impacts of having conductance-based synaptic coupling on synchronizability.  Chemical synapses are not strictly excitatory or inhibitory as the flow of current is dictated by the driving force, which leads to complex behaviours  and synchronizability regimes, where excitation can lead to desynchronization or inhibition can lead to synchronization. 

\subsection*{Acknowledgement} 
WN is supported by an NSERC Discovery Grant (DGECR/00334-2020) and a Canada Research Chair (CRC-2019-00416)
%\subsection*{Code Availability Statement} 
%Accompanying code can be found on modelDB \cite{modeldb}.  The link to the model is https://modeldb.science/2014997 with referee access code RFV114. 

\clearpage 
\section*{Appendix A}

\subsection*{The Master Stability Function}
The Master Stability Function (MSF) approach to resolving the stability of synchronous solutions was originally proposed in \cite{msf1}, with subsequent advances to the analysis of clustered systems in \cite{msf2}.  Briefly, the approach considers coupled non-linear systems of the following form:
\begin{eqnarray*}
\frac{d\bm x_i}{dt} = Q(\bm x_i) + \sum_{j=1}^N g_{ij} H(\bm x_j) , \quad i=1,2,\ldots N, \quad  \bm x_i \in \mathbb{R}^k ,  \label{msf1}
\end{eqnarray*}
with the constraint that 
\begin{eqnarray*}
\sum_{j=1}^N g_{ij} = 0, \forall i 
\end{eqnarray*}
and that the coupling matrix $\bm g$ be diagonalizable.  This is sometimes referred to as ``diffusive connectivity" 

One can express the system in (\ref{msf1}) with the direct product $\otimes$ or tensor operator as  
\begin{eqnarray*}
\frac{d\bm x}{dt} = \bm Q(\bm x) + \bm G \otimes \bm H(\bm x)
\end{eqnarray*}
where 
$$ \bm Q(\bm x) = [Q(\bm x_1),Q(\bm x_2),\ldots Q(\bm x_N)], \quad \bm H(\bm x) = [H(\bm x_1),H(\bm x_2)\ldots H(\bm x_N)].$$

Under these constraints, the synchronous solution $\bm x_s(t)$ is a solution to the differential equation 
\begin{eqnarray}
\frac{d\bm x_s}{dt} = Q(\bm x_s)\label{synchsol} 
\end{eqnarray}

Then, perturbations off of the synchronous solution are considered 
 \begin{eqnarray*}
\bm x_i(t) = \bm x_s(t) + \bm \epsilon_i(t).
\end{eqnarray*}
which yields the following block diagonal system 
\begin{eqnarray}
\frac{d\bm \epsilon}{dt} = \left[\bm I_N \otimes \frac{\partial Q}{\partial \bm x}  + \bm G \otimes \frac{\partial H}{\partial\bm x}\right]\bm \epsilon   \label{lin1} 
\end{eqnarray} 
The coupled non-autonomous system in (\ref{lin1}) determines the linear stability of the system (\ref{synchsol}) via computation of the Lyapunov exponents of the system.  In a master-stability function approach, the problem of computing these Lyapunov is simplified greatly by assuming that $\bm G$ is diagonalizable: 
\begin{eqnarray*}
\bm G = \bm P\bm D \bm P^{-1}  
\end{eqnarray*}
By applying the substitution $\bm \eta = \bm P^{-1} \bm \eta$, the system simplifies into a diagonalized block-system:  
\begin{eqnarray}
\frac{d\bm \eta_i}{dt} =  \left[\frac{\partial Q}{\partial \bm x}  +  \lambda_i \frac{\partial H}{\partial\bm x}\right]\bm \eta_i\label{block1}, \quad i = 1,2,\ldots N
\end{eqnarray} 
where $\lambda_i$ is an eigenvalue of the matrix $\bm G$.  The next step in a MSF function approach is to compute the Lyapunov exponents of equation \ref{block1} over a mesh in the complex eigenvalue space of $\lambda = a + b i $.  Thus, one computes the mesh of lyapunov exponents, $\mu$ as a function of the eigenvalues $\lambda$: 
\begin{eqnarray}
\frac{d\bm \eta}{dt} =  \left[\frac{\partial Q}{\partial \bm x}  +  \lambda \frac{\partial H}{\partial\bm x}\right]\bm \eta, \rightarrow \mu(\lambda)  \label{block2}
\end{eqnarray} 
 The Lyapunov exponents $\mu(\lambda)$ can be computed with established methods for their approximation (e.g. \cite{wolf}). 

The maximum Lyapunov exponent of the block \ref{block1} as function of $\lambda$ is the Master Stability Function (MSF).  With the mesh computed, one can use any diagonalizable connection matrix and simply "look up" the value of the maximum Lyapunov exponent as a function of $\lambda$ with $\mu(\lambda)$, the MSF.    In this work, we refer to the final diagonalized block structure in (\ref{block2}) as the MSF equations, as they are necessary for the numerical approximation of $\mu(\lambda)$.

\subsection*{\textbf{Supplementary Material } }

The parameters for the Morris-Lecar neuron model under Hodgkin Class I and Hodgkin Class 2 regimes are shown in Table \ref{table1}, along with the synaptic parameters. 
\begin{table}[htp!]
\centering
\begin{tabular}{ |c |c| c |}
\hline
 Parameter & Value & Units  \\
\hline 
 $C$ & 20 & pF \\  
\hline 
 $g_L$ & 2 & nS \\  
\hline 
 $E_L$ & 60 & mV \\  
\hline 
 $g_K$ & 8 & nS \\  
\hline 
 $E_K$ & -84 & mV \\  
\hline 
 $g_{Ca}$ & 4 & nS \\  
\hline 
 $E_{Ca}$ & 120 & mV \\  
\hline 
 $V_1$ & -1.2 & mV \\  
\hline
 $V_2$ & 18 & mV \\  
\hline
 $V_3$ & 12 (Class I), 2 (Class II)  & mV \\  
\hline
 $V_4$ & 17.4 (Class I), 30 (Class II) & mV \\  
\hline    
 $\phi$ & 0.067 (Class I), 0.04 (Class II)& mV \\  
\hline 
 $I$ & 50 (Class I), 115 (Class II)& pA \\  
\hline 
 $a_r$ & 1.1 & ms$^{-1}$ \\  
\hline 
$a_d$ & 0.19 & ms$^{-1}$ \\  
\hline 
$T_{max}$ & 1& unitless \\
\hline
$K_p$ & 5 & mV \\
\hline
$V_t$ & 2 & mV \\
\hline
\end{tabular}
\caption{Parameter values used for the Morris-Lecar neuron model, unless otherwise specified in a figure} \label{table1}
\end{table}

\subsection*{\textbf{Numerical Integration and Computation of the Lyapunov Exponents}}

All ODEs for the Morris-Lecar system(s) under consideration were integrated with the MATLAB2023a function \emph{ode45}.  The 'RelTol' (relative error tolerance) and 'AbsTol' absolute error tolerance parametes were set to $10^{-14}$ for all direct integration of the Morris-Lecar network equations.  The Lypaunov exponents were computed with the algorithm in \cite{wolf} with code modified from \cite{w3,w4}.


\clearpage

\begin{thebibliography}{10}
\expandafter\ifx\csname url\endcsname\relax
  \def\url#1{\texttt{#1}}\fi
\expandafter\ifx\csname urlprefix\endcsname\relax\def\urlprefix{URL }\fi
\providecommand{\bibinfo}[2]{#2}
\providecommand{\eprint}[2][]{\url{#2}}

\bibitem{Buz1}
\bibinfo{author}{Buzs{\'a}ki, G.}
\newblock \bibinfo{title}{Hippocampal sharp wave-ripple: A cognitive biomarker
  for episodic memory and planning}.
\newblock \emph{\bibinfo{journal}{Hippocampus}} \textbf{\bibinfo{volume}{25}},
  \bibinfo{pages}{1073--1188} (\bibinfo{year}{2015}).

\bibitem{Wilson1}
\bibinfo{author}{Lee, A.~K.} \& \bibinfo{author}{Wilson, M.~A.}
\newblock \bibinfo{title}{Memory of sequential experience in the hippocampus
  during slow wave sleep}.
\newblock \emph{\bibinfo{journal}{Neuron}} \textbf{\bibinfo{volume}{36}},
  \bibinfo{pages}{1183--1194} (\bibinfo{year}{2002}).

\bibitem{Wilson2}
\bibinfo{author}{Wilson, M.~A.} \& \bibinfo{author}{McNaughton, B.~L.}
\newblock \bibinfo{title}{Reactivation of hippocampal ensemble memories during
  sleep}.
\newblock \emph{\bibinfo{journal}{Science}} \textbf{\bibinfo{volume}{265}},
  \bibinfo{pages}{676--679} (\bibinfo{year}{1994}).

\bibitem{msf1}
\bibinfo{author}{Pecora, L.~M.} \& \bibinfo{author}{Carroll, T.~L.}
\newblock \bibinfo{title}{Master stability functions for synchronized coupled
  systems}.
\newblock \emph{\bibinfo{journal}{Physical review letters}}
  \textbf{\bibinfo{volume}{80}}, \bibinfo{pages}{2109} (\bibinfo{year}{1998}).

\bibitem{msf2}
\bibinfo{author}{Pecora, L.~M.}, \bibinfo{author}{Sorrentino, F.},
  \bibinfo{author}{Hagerstrom, A.~M.}, \bibinfo{author}{Murphy, T.~E.} \&
  \bibinfo{author}{Roy, R.}
\newblock \bibinfo{title}{Cluster synchronization and isolated
  desynchronization in complex networks with symmetries}.
\newblock \emph{\bibinfo{journal}{Nature communications}}
  \textbf{\bibinfo{volume}{5}}, \bibinfo{pages}{4079} (\bibinfo{year}{2014}).

\bibitem{coombesmsf1}
\bibinfo{author}{Coombes, S.} \& \bibinfo{author}{Thul, R.}
\newblock \bibinfo{title}{Synchrony in networks of coupled non-smooth dynamical
  systems: Extending the master stability function} .

\bibitem{coombesmsf2}
\bibinfo{author}{Lai, Y.~M.}, \bibinfo{author}{Thul, R.} \&
  \bibinfo{author}{Coombes, S.}
\newblock \bibinfo{title}{Analysis of networks where discontinuities and
  nonsmooth dynamics collide: understanding synchrony}.
\newblock \emph{\bibinfo{journal}{The European Physical Journal Special
  Topics}} \textbf{\bibinfo{volume}{227}}, \bibinfo{pages}{1251--1265}
  (\bibinfo{year}{2018}).

\bibitem{coombesmsf3}
\bibinfo{author}{Coombes, S.} \& \bibinfo{author}{Wedgwood, K.~C.}
\newblock \emph{\bibinfo{title}{Neurodynamics: An Applied Mathematics
  Perspective}}, vol.~\bibinfo{volume}{75} (\bibinfo{publisher}{Springer
  Nature}, \bibinfo{year}{2023}).

\bibitem{Izhikevich1}
\bibinfo{author}{Izhikevich, E.~M.}
\newblock \bibinfo{title}{Simple model of spiking neurons}.
\newblock \emph{\bibinfo{journal}{IEEE Transactions on neural networks}}
  \textbf{\bibinfo{volume}{14}}, \bibinfo{pages}{1569--1572}
  (\bibinfo{year}{2003}).

\bibitem{Izhikevich2}
\bibinfo{author}{Izhikevich, E.~M.}
\newblock \emph{\bibinfo{title}{Dynamical systems in neuroscience}}
  (\bibinfo{publisher}{MIT press}, \bibinfo{year}{2007}).

\bibitem{Touboul1}
\bibinfo{author}{Touboul, J.}
\newblock \bibinfo{title}{Bifurcation analysis of a general class of nonlinear
  integrate-and-fire neurons}.
\newblock \emph{\bibinfo{journal}{SIAM Journal on Applied Mathematics}}
  \textbf{\bibinfo{volume}{68}}, \bibinfo{pages}{1045--1079}
  (\bibinfo{year}{2008}).

\bibitem{Adex1}
\bibinfo{author}{Brette, R.} \& \bibinfo{author}{Gerstner, W.}
\newblock \bibinfo{title}{Adaptive exponential integrate-and-fire model as an
  effective description of neuronal activity}.
\newblock \emph{\bibinfo{journal}{Journal of neurophysiology}}
  \textbf{\bibinfo{volume}{94}}, \bibinfo{pages}{3637--3642}
  (\bibinfo{year}{2005}).

\bibitem{Adex2}
\bibinfo{author}{Naud, R.}, \bibinfo{author}{Marcille, N.},
  \bibinfo{author}{Clopath, C.} \& \bibinfo{author}{Gerstner, W.}
\newblock \bibinfo{title}{Firing patterns in the adaptive exponential
  integrate-and-fire model}.
\newblock \emph{\bibinfo{journal}{Biological cybernetics}}
  \textbf{\bibinfo{volume}{99}}, \bibinfo{pages}{335--347}
  (\bibinfo{year}{2008}).

\bibitem{coombesmsf4}
\bibinfo{author}{Ashwin, P.}, \bibinfo{author}{Coombes, S.} \&
  \bibinfo{author}{Nicks, R.}
\newblock \bibinfo{title}{Mathematical frameworks for oscillatory network
  dynamics in neuroscience}.
\newblock \emph{\bibinfo{journal}{The Journal of Mathematical Neuroscience}}
  \textbf{\bibinfo{volume}{6}}, \bibinfo{pages}{1--92} (\bibinfo{year}{2016}).

\bibitem{destexhe}
\bibinfo{author}{Destexhe, A.}, \bibinfo{author}{Mainen, Z.~F.},
  \bibinfo{author}{Sejnowski, T.~J.} \emph{et~al.}
\newblock \bibinfo{title}{Kinetic models of synaptic transmission}.
\newblock \emph{\bibinfo{journal}{Methods in neuronal modeling}}
  \textbf{\bibinfo{volume}{2}}, \bibinfo{pages}{1--25} (\bibinfo{year}{1998}).

\bibitem{gj1}
\bibinfo{author}{Lehnert, J.}, \bibinfo{author}{Dahms, T.},
  \bibinfo{author}{H{\"o}vel, P.} \& \bibinfo{author}{Sch{\"o}ll, E.}
\newblock \bibinfo{title}{Loss of synchronization in complex neuronal networks
  with delay}.
\newblock \emph{\bibinfo{journal}{Europhysics Letters}}
  \textbf{\bibinfo{volume}{96}}, \bibinfo{pages}{60013} (\bibinfo{year}{2011}).

\bibitem{gj2}
\bibinfo{author}{Bonacini, E.} \emph{et~al.}
\newblock \bibinfo{title}{How single node dynamics enhances synchronization in
  neural networks with electrical coupling}.
\newblock \emph{\bibinfo{journal}{Chaos, Solitons \& Fractals}}
  \textbf{\bibinfo{volume}{85}}, \bibinfo{pages}{32--43}
  (\bibinfo{year}{2016}).

\bibitem{gj3}
\bibinfo{author}{Jalili, M.}
\newblock \bibinfo{title}{Enhancing synchronizability of diffusively coupled
  dynamical networks: A survey}.
\newblock \emph{\bibinfo{journal}{IEEE transactions on neural networks and
  learning systems}} \textbf{\bibinfo{volume}{24}}, \bibinfo{pages}{1009--1022}
  (\bibinfo{year}{2013}).

\bibitem{gj4}
\bibinfo{author}{Belykh, I.}, \bibinfo{author}{De~Lange, E.} \&
  \bibinfo{author}{Hasler, M.}
\newblock \bibinfo{title}{Synchronization of bursting neurons: What matters in
  the network topology}.
\newblock \emph{\bibinfo{journal}{Physical review letters}}
  \textbf{\bibinfo{volume}{94}}, \bibinfo{pages}{188101}
  (\bibinfo{year}{2005}).

\bibitem{gj5}
\bibinfo{author}{Neefs, P.}, \bibinfo{author}{Steur, E.} \&
  \bibinfo{author}{Nijmeijer, H.}
\newblock \bibinfo{title}{Network complexity and synchronous behavior—an
  experimental approach}.
\newblock \emph{\bibinfo{journal}{International Journal of Neural Systems}}
  \textbf{\bibinfo{volume}{20}}, \bibinfo{pages}{233--247}
  (\bibinfo{year}{2010}).

\bibitem{gj6}
\bibinfo{author}{Keplinger, K.} \& \bibinfo{author}{Wackerbauer, R.}
\newblock \bibinfo{title}{Transient spatiotemporal chaos in the morris-lecar
  neuronal ring network}.
\newblock \emph{\bibinfo{journal}{Chaos: An Interdisciplinary Journal of
  Nonlinear Science}} \textbf{\bibinfo{volume}{24}} (\bibinfo{year}{2014}).

\bibitem{gj7}
\bibinfo{author}{Rossoni, E.}, \bibinfo{author}{Chen, Y.},
  \bibinfo{author}{Ding, M.} \& \bibinfo{author}{Feng, J.}
\newblock \bibinfo{title}{Stability of synchronous oscillations in a system of
  hodgkin-huxley neurons with delayed diffusive and pulsed coupling}.
\newblock \emph{\bibinfo{journal}{Physical Review E}}
  \textbf{\bibinfo{volume}{71}}, \bibinfo{pages}{061904}
  (\bibinfo{year}{2005}).

\bibitem{ml1}
\bibinfo{author}{Morris, C.} \& \bibinfo{author}{Lecar, H.}
\newblock \bibinfo{title}{Voltage oscillations in the barnacle giant muscle
  fiber}.
\newblock \emph{\bibinfo{journal}{Biophysical journal}}
  \textbf{\bibinfo{volume}{35}}, \bibinfo{pages}{193--213}
  (\bibinfo{year}{1981}).

\bibitem{bardtext}
\bibinfo{author}{Ermentrout, B.} \& \bibinfo{author}{Terman, D.~H.}
\newblock \emph{\bibinfo{title}{Mathematical foundations of neuroscience}},
  vol.~\bibinfo{volume}{35} (\bibinfo{publisher}{Springer},
  \bibinfo{year}{2010}).

\bibitem{msf3}
\bibinfo{author}{Huang, L.}, \bibinfo{author}{Chen, Q.}, \bibinfo{author}{Lai,
  Y.-C.} \& \bibinfo{author}{Pecora, L.~M.}
\newblock \bibinfo{title}{Generic behavior of master-stability functions in
  coupled nonlinear dynamical systems}.
\newblock \emph{\bibinfo{journal}{Physical Review E}}
  \textbf{\bibinfo{volume}{80}}, \bibinfo{pages}{036204}
  (\bibinfo{year}{2009}).

\bibitem{cjgj1}
\bibinfo{author}{Jhou, F.-J.}, \bibinfo{author}{Juang, J.} \&
  \bibinfo{author}{Liang, Y.-H.}
\newblock \bibinfo{title}{Multistate and multistage synchronization of
  hindmarsh-rose neurons with excitatory chemical and electrical synapses}.
\newblock \emph{\bibinfo{journal}{IEEE Transactions on Circuits and Systems I:
  Regular Papers}} \textbf{\bibinfo{volume}{59}}, \bibinfo{pages}{1335--1347}
  (\bibinfo{year}{2012}).

\bibitem{cjgj2}
\bibinfo{author}{Aristides, R.~P.} \& \bibinfo{author}{Cerdeira, H.~A.}
\newblock \bibinfo{title}{Master stability functions of networks of izhikevich
  neurons}.
\newblock \emph{\bibinfo{journal}{arXiv preprint arXiv:2303.13921}}
  (\bibinfo{year}{2023}).

\bibitem{w1}
\bibinfo{author}{Checco, P.}, \bibinfo{author}{Righero, M.},
  \bibinfo{author}{Biey, M.} \& \bibinfo{author}{Kocarev, L.}
\newblock \bibinfo{title}{Synchronization in networks of hindmarsh--rose
  neurons}.
\newblock \emph{\bibinfo{journal}{IEEE Transactions on Circuits and Systems II:
  Express Briefs}} \textbf{\bibinfo{volume}{55}}, \bibinfo{pages}{1274--1278}
  (\bibinfo{year}{2008}).

\bibitem{w2}
\bibinfo{author}{Nicola, W.}, \bibinfo{author}{Hellyer, P.~J.},
  \bibinfo{author}{Campbell, S.~A.} \& \bibinfo{author}{Clopath, C.}
\newblock \bibinfo{title}{Chaos in homeostatically regulated neural systems}.
\newblock \emph{\bibinfo{journal}{Chaos: An Interdisciplinary Journal of
  Nonlinear Science}} \textbf{\bibinfo{volume}{28}} (\bibinfo{year}{2018}).

\bibitem{w3}
\bibinfo{author}{Nicola, W.} \& \bibinfo{author}{Campbell, S.~A.}
\newblock \bibinfo{title}{Normalized connectomes show increased
  synchronizability with age through their second largest eigenvalue}.
\newblock \emph{\bibinfo{journal}{SIAM Journal on Applied Dynamical Systems}}
  \textbf{\bibinfo{volume}{20}}, \bibinfo{pages}{1158--1176}
  (\bibinfo{year}{2021}).

\bibitem{w4}
\bibinfo{author}{Al-Darabsah, I.}, \bibinfo{author}{Chen, L.},
  \bibinfo{author}{Nicola, W.} \& \bibinfo{author}{Campbell, S.~A.}
\newblock \bibinfo{title}{The impact of small time delays on the onset of
  oscillations and synchrony in brain networks}.
\newblock \emph{\bibinfo{journal}{Frontiers in Systems Neuroscience}}
  \textbf{\bibinfo{volume}{15}}, \bibinfo{pages}{688517}
  (\bibinfo{year}{2021}).

\bibitem{inhibs1}
\bibinfo{author}{Van~Vreeswijk, C.}, \bibinfo{author}{Abbott, L.} \&
  \bibinfo{author}{Bard~Ermentrout, G.}
\newblock \bibinfo{title}{When inhibition not excitation synchronizes neural
  firing}.
\newblock \emph{\bibinfo{journal}{Journal of computational neuroscience}}
  \textbf{\bibinfo{volume}{1}}, \bibinfo{pages}{313--321}
  (\bibinfo{year}{1994}).

\bibitem{inhibs4}
\bibinfo{author}{White, J.~A.}, \bibinfo{author}{Chow, C.~C.},
  \bibinfo{author}{Rit, J.}, \bibinfo{author}{Soto-Trevi{\~n}o, C.} \&
  \bibinfo{author}{Kopell, N.}
\newblock \bibinfo{title}{Synchronization and oscillatory dynamics in
  heterogeneous, mutually inhibited neurons}.
\newblock \emph{\bibinfo{journal}{Journal of computational neuroscience}}
  \textbf{\bibinfo{volume}{5}}, \bibinfo{pages}{5--16} (\bibinfo{year}{1998}).

\bibitem{inhibs2}
\bibinfo{author}{Elson, R.~C.}, \bibinfo{author}{Selverston, A.~I.},
  \bibinfo{author}{Abarbanel, H.~D.} \& \bibinfo{author}{Rabinovich, M.~I.}
\newblock \bibinfo{title}{Inhibitory synchronization of bursting in biological
  neurons: dependence on synaptic time constant}.
\newblock \emph{\bibinfo{journal}{Journal of Neurophysiology}}
  \textbf{\bibinfo{volume}{88}}, \bibinfo{pages}{1166--1176}
  (\bibinfo{year}{2002}).

\bibitem{inhibs3}
\bibinfo{author}{Khazipov, R.}
\newblock \bibinfo{title}{Gabaergic synchronization in epilepsy}.
\newblock \emph{\bibinfo{journal}{Cold Spring Harbor perspectives in medicine}}
  \textbf{\bibinfo{volume}{6}} (\bibinfo{year}{2016}).

\bibitem{modeldb}
\bibinfo{author}{McDougal, R.~A.} \emph{et~al.}
\newblock \bibinfo{title}{Twenty years of modeldb and beyond: building
  essential modeling tools for the future of neuroscience}.
\newblock \emph{\bibinfo{journal}{Journal of computational neuroscience}}
  \textbf{\bibinfo{volume}{42}}, \bibinfo{pages}{1--10} (\bibinfo{year}{2017}).

\bibitem{wolf}
\bibinfo{author}{Wolf, A.}, \bibinfo{author}{Swift, J.~B.},
  \bibinfo{author}{Swinney, H.~L.} \& \bibinfo{author}{Vastano, J.~A.}
\newblock \bibinfo{title}{Determining lyapunov exponents from a time series}.
\newblock \emph{\bibinfo{journal}{Physica D: nonlinear phenomena}}
  \textbf{\bibinfo{volume}{16}}, \bibinfo{pages}{285--317}
  (\bibinfo{year}{1985}).

\end{thebibliography}
\end{document}